\documentclass[english,prl,twocolumn,amsmath,amssymb,showpacs]{revtex4}
\usepackage[T1]{fontenc}
\usepackage[latin1]{inputenc}
\usepackage{amsmath}
\usepackage{graphicx}
\usepackage{amssymb}

\makeatletter

\usepackage{graphicx}% Include figure files
\usepackage{bm}% bold math

\usepackage{babel}
\makeatother
\begin{document}

\preprint{v. 2.05}

\title{Classical chaos with Bose-Einstein condensates in tilted optical
lattices }

\author{Quentin Thommen}

\author{Jean Claude Garreau}

\author{V\'{e}ronique Zehnlé}

\affiliation{Laboratoire de Physique des Lasers, Atomes et Mol\'{e}cules, Centre
d'Etudes et de Recherches Laser et Applications, Universit\'{e} des
Sciences et Technologies de Lille, F-59655 Villeneuve d'Ascq Cedex,
France}

\homepage{http://www.phlam.univ-lille1.fr/atfr/cq}

\date{\today{}}

\begin{abstract}
A widely accepted definition of {}``quantum chaos'' is {}``the
behavior of a quantum system whose \emph{classical} \emph{limit is
chaotic}''. The dynamics of quantum-chaotic systems is nevertheless
very different from that of their classical counterparts. A fundamental
reason for that is the linearity of Schrödinger equation. In this
paper, we study the quantum dynamics of an ultra-cold quantum degenerate
gas in a tilted optical lattice and show that it displays features
very close to \emph{classical} chaos. We show that its phase space
is organized according to the Kolmogorov-Arnold-Moser theorem.
\end{abstract}


\maketitle
The quantum dynamics of systems presenting chaotic behavior in the
classical limit has been a widely studied subject in recent years,
boosted in particular by experiments using laser-cooled atoms \cite{Raizen_LDynFirst_PRL94,Christ_LDynNoise_PRL98,Burnett_AccModes_PRL99,Bicolor_PRL00,SubFourier_PRL02}.
An atomic-scale analog of the kicked rotor is realized by placing
laser-cooled atoms in a pulsed standing laser wave, a system that
displays a well-known \emph{signature} of quantum chaos, the \emph{dynamical
localization} \cite{Casati_LocDynFirst_LNP79}, present in time-periodic
systems, which consists in the suppression of the classical chaotic
diffusion by quantum interferences. Another well-known signature of
quantum chaos, present in time-independent systems, is the fact that
the distribution of energy levels takes the shape of a Wigner distribution,
a behavior experimentally evidenced for example in the energy spectrum
of Rydberg atoms in intense magnetic fields. This signature clearly
cannot have a classical counterpart. 

Quite generally, signatures of quantum chaos have no direct relation
to the corresponding classical dynamics. There are fundamental reasons
for this. One is that the notion of phase-space orbit, fundamental
in classical dynamics, cannot be easily translated in the quantum
world, due to the uncertainty principle. The initial conditions of
a quantum system do not correspond to a single point in the usual
phase space $(q_{i},p_{i})$, as $\Delta q_{i}\Delta p_{i}\ge \hbar /2$.
One can however project the state of the system in a basis, and define
a generalized phase space formed by the set of amplitudes; if the
basis is adequately chosen, the initial condition is represented by
a point in such generalized phase space. Even then, sensitivity to
initial conditions in not present, for the more fundamental reason
that the Schrödinger equation is linear. Nonlinearity arises in classical
physics because the dynamical variables $(q_{i},p_{i})$ are also
parameters of the force. In the case of the Schrödinger equation,
the dynamical variable is the wave function, whereas the potential
depends on $(q_{i},p_{i})$, and eventually on $t$, but not on the
wave function. 

A Bose-Einstein condensate (BEC) is a quantum objects described by
a \emph{nonlinear} quantum evolution equation, the Gross-Pitaevskii
equation (GPE). GPE has proved to be able to describe the BEC dynamics
with a reasonable precision on a wide range of situations \cite{Sringari_BECRevTh_RMP99}.
GPE describes the condensate as whole (neglecting the non-condensed
fraction), and includes a nonlinear term representing particle-particle
interactions. It is valid for temperatures low enough compared to
the critical temperature. If we consider a cigar-shaped BEC whose
transverse length $L$ is much smaller than the longitudinal length,
Gross-Pitaevskii equation can be reduced to one-dimension, and reads:\begin{equation}
i\hbar \frac{\partial \psi }{\partial t}=\left(H_{0}+U_{0}\left|\psi \right|²\right)\psi \label{eq:GPE}\end{equation}
where $H_{0}$ is the one-particle Hamiltonian, and $U_{0}$ a coupling
constant describing the interaction among particles. $U_{0}$ is related
to the two-particle $s$-wave scattering length $a_{s}$ by $U_{0}=4\pi \hbar ^{2}a_{s}N/(L^{2}M)$,
where $N$ is the total number of atoms, $M$ the mass of an atom,
and the condensate wave function is normalized to unity. Contrary
to the one-particle Schrödinger equation, this equation is nonlinear
in the dynamical variable $\psi $. Note that, as $U_{0}\propto N$,
the GPE nonlinearity has a \emph{collective} character: modeling in
an analogous way many-body interactions in an assembly of $N$ thermal
atoms would produce a much smaller nonlinear term. 

The dynamics of one dimension, tilted, periodic potentials has been
often considered in the recent literature, both for free atoms \cite{Salomon_BlochOsc_PRL96,Korsch_LifetimeWS_PRL99,Nienhuis_CoherentDyn_PRA01,WannierStark_PRA02},
and for BECs \cite{Kasevich_BECTiltedLattice_Science98,Arimondo_BECBlochOsc_PRL01,Trombettoni_SolitonsBEC_PRL01}.
A tilted optical potential\begin{equation}
V=V_{0}\cos (\frac{2\pi x}{d})+Fx\label{eq:TiltedPot}\end{equation}
is generated by superposing two counterpropagating laser waves, one
of frequency $\omega _{L}=k_{L}c=2\pi c/\lambda _{L}$ and the other
of frequency $\omega _{L}(1+\gamma t)$. This produces a standing
wave of step $d=\lambda _{L}/2$ whose nodes are accelerated with
an acceleration $c\gamma /2$. In the (non-inertial) reference frame
in which the nodes of the standing wave are at rest, an inertial force
$F=-Mc\gamma /2$ appears, producing a {}``tilted'' potential. In
such potential, quantum dynamics consists in spatial oscillations
over many steps of the potential, known as Bloch oscillations, as
recently experimentally observed with both individual atoms \cite{Salomon_BlochOsc_PRL96}
and (weakly interacting -- low $U_{0}$) Bose-Einstein condensates
\cite{Arimondo_BECBlochOsc_PRL01}.

\begin{figure}[htbp]
\includegraphics[  clip,
  angle=270,
  origin=lB]{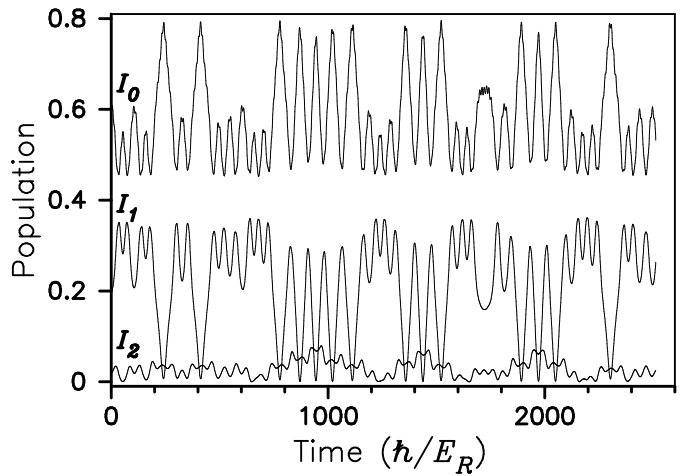}

\caption{\label{fig:EvolutionGPE}Erratic evolution of the populations. The
BEC is initially prepared on twelve neighbor WS states, and the figure
displays the evolution of the populations $I_{n}=\left|\langle \varphi _{n}|\psi \rangle \right|^{2}$
of the three central (most populated) wells. The result, obtained
by numerical integration of the GPE, clearly show an erratic behavior.
Parameters are $V_{0}=5$, $F=0.25$, and $g=0.25$.}
\end{figure}

We shall consider, in what follows, the particular form of the GPE
given by\begin{equation}
i\frac{\partial \psi (x,t)}{\partial t}=\left[-\frac{1}{2m}\frac{\partial ^{2}}{\partial x^{2}}+V_{0}\cos \left(2\pi x\right)+Fx+g\left|\psi \right|^{2}\right]\psi ,\label{eq:GPE_Tilted}\end{equation}
where we introduced normalized units in which space is measured in
units of the potential step $d$, energy in units of the recoil energy
$E_{R}=\hbar ^{2}k_{L}^{2}/2M$, time in units of $\hbar /E_{R}$,
force in units of $2E_{R}/\lambda _{L}$, $m=\pi ^{2}/2$, and $g=8a_{s}N/(L^{2}k_{L})=(2/\lambda _{L})(U_{0}/E_{R})$
is the (1D) nonlinear parameter ($\hbar =1$ in such units). The eigenstates
of the linear part of the Hamiltonian\begin{equation}
H_{0}=-\frac{1}{2m}\frac{\partial ^{2}}{\partial x^{2}}+V_{0}\cos \left(2\pi x\right)+Fx,\label{eq:H0}\end{equation}
 are the so-called Wannier-Stark (WS) states %
\footnote{This is rigorously true if the whole system is enclosed in a bounding
box of dimensions much larger than $d$; otherwise, WS states must
be considered as metastable states (resonances).%
}. We assume that the depth of the potential $V_{0}$ is large enough
to support well localized WS states, and that the dynamics can be
described by the lowest-energy WS state in each well \cite{WannierStark_PRA02}.
We note by $\varphi _{n}$ the lowest state mainly localized in the
$n^{th}$-well. The invariance of WS states under translations by
an integer number of potential steps, $\varphi _{n}(x)=\varphi _{0}(x-n)$,
is related to shifts of the eigenenergies given by $E_{n}=E_{0}+nF$
($E_{0}$ is an energy constant that can be eliminated), forming the
so-called Wannier-Stark ladder of equally spaced levels %
\footnote{These symmetries are approximate for WS states in a bounding box,
but they apply with a good precision to states far from the bounding
box.%
}, that define the Bloch (or Bohr) frequency of the system $\omega _{B}=Fd/\hbar $.
Fig. \ref{fig:EvolutionGPE} displays an example of evolution of a
BEC in a tilted potential, and the presence of an erratic behavior
suggests the existence of \emph{chaos} in a \emph{classical} sense. 

\begin{figure}
\includegraphics[  angle=270,
  origin=lB]{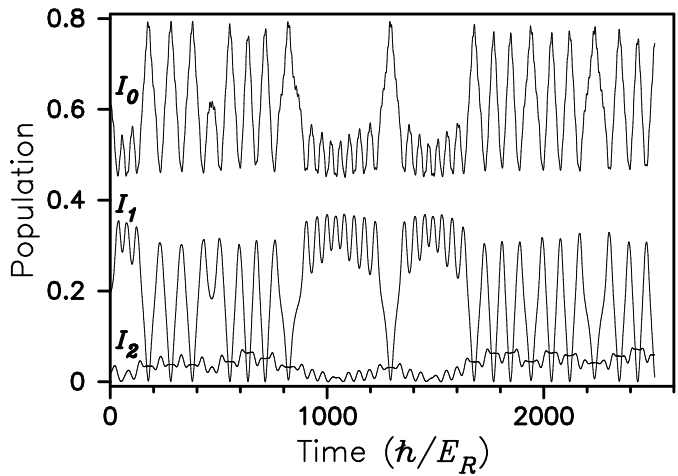}

\caption{\label{fig:EvolutionDiscrete}Same as Fig. \ref{fig:EvolutionGPE},
but the calculation was made using the discrete model of Eq. (\ref{eq:EqMotion}).
The dynamical behavior is the same as in Fig. \ref{fig:EvolutionGPE}.}
\end{figure}

In order to get a simpler description and a better understanding of
the BEC dynamics, we now introduce a model obtained by projecting
the wave function $\psi $ over the WS states: \begin{equation}
\psi (x,t)=\sum \limits _{n}c_{n}(t)\varphi _{n}(x)\label{eq:superposition}\end{equation}
 where the $c_{n}(t)\equiv \sqrt{I_{n}}e^{i\theta _{n}}$ are complex
amplitudes. This development is justified provided the $g\lesssim 1$
\footnote{Numerical integration of Eq. (\ref{eq:GPE_Tilted}) confirms the completeness
of Eq. (\ref{eq:superposition}) for values of $g$ as high as $g=1$.
For typical evolution times, $\sum _{n}\left|c_{n}(t)\right|^{2}$
remains very close to one: the loss of {}``completeness'' is of
a few percent.%
}, which corresponds, taking typical values for the parameters, to
$N\le 7\times 10^{4}$. Reporting Eq. (\ref{eq:superposition}) in
Eq. (\ref{eq:GPE_Tilted}) gives the evolution equation:\begin{equation}
i\dot{c}_{n}=nFc_{n}+g\sum _{klm}\chi _{klm}^{n}c_{k}^{*}c_{l}c_{m}.\label{eq:GP1}\end{equation}
where the $\chi _{klm}^{n}$ are defined (choosing phases such that
$\varphi _{n}$ is a real function) by:\begin{equation}
\chi _{klm}^{n}=\int _{-\infty }^{\infty }\varphi _{k}(x)\varphi _{l}(x)\varphi _{m}(x)\varphi _{n}(x)dx.\label{eq:couplage}\end{equation}
Due to invariance under discrete spatial translations of the WS states
one easily shows that $\chi _{k-n,l-n,m-n}^{n}=\chi _{klm}^{0}\equiv \chi _{klm}$,
and thus:\[
i\dot{c}_{n}=nFc_{n}+g\sum _{klm}\chi _{klm}c_{n+k}^{*}c_{n+l}c_{n+m}.\]
As WS are essentially localized in a well, we can keep only couplings
between nearest neighbors, that is, involving $\chi _{000}$, $\chi _{00\pm 1}$
\footnote{\label{foot:Chi}For the typical parameters we are dealing with, $V_{0}=5$
and $F=0.25$, one obtains: $\chi _{000}\approx 1.99$, $\chi _{001}\approx -0.16$,
and $\chi _{00-1}\approx 0.15$. Other couplings are much smaller.%
}. The equation of motion then simplifies to:\begin{eqnarray}
i\dot{c_{n}} & = & nFc_{n}+g\chi _{000}c_{n}\left|c_{n}\right|^{2}+\nonumber \\
 &  & g\left(\chi _{00-1}c_{n-1}^{*}+\chi _{001}c_{n+1}^{*}\right)c_{n}^{2}+\nonumber \\
 &  & g\left(2\chi _{00-1}c_{n-1}+2\chi _{001}c_{n+1}\right)\left|c_{n}\right|^{2}+\nonumber \\
 &  & g\left(\chi _{001}\left|c_{n-1}\right|^{2}c_{n-1}+\chi _{00-1}\left|c_{n+1}\right|^{2}c_{n+1}\right)\label{eq:EqMotion}
\end{eqnarray}
where we used the invariance of the $\chi _{klm}$ coefficients under
any permutation of indexes. Fig. \ref{fig:EvolutionDiscrete} displays
the evolution obtained from the model, for the same parameters and
the same initial conditions as in Fig. \ref{fig:EvolutionGPE}, and
one observes the same dynamical behavior. The model is thus able to
reproduce the dynamics of the GPE.

It is useful to write evolution equations for the population $I_{n}=\left|c_{n}\right|²$
and the phase $\theta _{n}$:

\begin{eqnarray}
\dot{I_{n}} & = & 2\varepsilon \sigma _{g}\sqrt{I_{n}}\left[\sqrt{I_{n+1}}\left(I_{n+1}+\beta I_{n}\right)\sin \left(\theta _{n+1}-\theta _{n}\right)+\right.\nonumber \\
 &  & \left.\sqrt{I_{n-1}}\left(I_{n}+\beta I_{n-1}\right)\sin \left(\theta _{n-1}-\theta _{n}\right)\right]\label{eq:Inevol}\\
\dot{\theta _{n}} & = & -nF-\sigma _{g}I_{n}-\nonumber \\
 &  & \frac{\varepsilon \sigma _{g}}{\sqrt{I_{n}}}\left[\sqrt{I_{n+1}}\left(I_{n+1}+3\beta I_{n}\right)\cos \left(\theta _{n+1}-\theta _{n}\right)-\right.\nonumber \\
 &  & \left.\sqrt{I_{n-1}}\left(3I_{n}+\beta I_{n-1}\right)\cos \left(\theta _{n-1}-\theta _{n}\right)\right]\label{eq:thetaevol}
\end{eqnarray}
with $\sigma _{g}=g/\left|g\right|$, $\beta =\chi _{001}/\chi _{00-1}\approx -1$,
and we have rescaled the time variable as $t\rightarrow t/(\left|g\right|\chi _{000})$
and the force as $F\rightarrow F/(\left|g\right|\chi _{000})$. Parameter
$\varepsilon =\chi _{00-1}/\chi _{000}$ is a small since $\chi _{00-1}/\chi _{000}\ll 1$
\ref{foot:Chi}. If we neglect inter-well particle interactions, i.e.
for $\varepsilon =0$, all $I_{n}=I_{n}(0)$ are constants of motion,
and $\theta _{n}=\theta _{n}(0)+\omega _{n}t$ increases linearly
with time with frequency $\omega _{n}=-nF-\sigma _{g}I_{n}$. The
variables $(I_{n},\theta _{n})$ have then an {}``action-angle''
structure and the system is \emph{integrable}. The trajectories lay
on tori defined by the values of the constants of motion \cite{GutzwillerChaos}.
In usual units the frequencies $\omega _{n}$ are\begin{equation}
\omega _{n}=n\omega _{B}+\frac{U_{0}\chi _{000}}{\hbar }I_{n},\label{eq:Freq}\end{equation}
where the second term on the right-hand side represents a {}``frequency
pulling'' effect due to the nonlinearity. Moreover, one can verify
that Eqs. (\ref{eq:Inevol}) and (\ref{eq:thetaevol}) can be obtained
by the usual prescription, $\dot{I_{n}}=\frac{\partial H}{\partial \theta _{n}}$
and $\dot{\theta _{n}}=-\frac{\partial H}{\partial I_{n}}$ from the
Hamiltonian given by\begin{eqnarray*}
H & = & \sum _{n}\left[nFI_{n}+\frac{\sigma _{g}}{2}I_{n}^{2}+\right.\\
 &  & \left.2\sigma _{g}\varepsilon \sqrt{I_{n}I_{n+1}}\left(\beta I_{n}+I_{n+1}\right)\cos \left(\theta _{n+1}-\theta _{n}\right)\right]\\
 & = & H_{0}(I)+\varepsilon \sum _{n}\left[V_{n}(I)\cos \left(\theta _{n+1}-\theta _{n}\right)\right].
\end{eqnarray*}
 The Hamiltonian is the sum of an integrable part and an non-integrable
{}``perturbation''. This form strongly evokes the Kolmogorov-Arnold-Moser
(KAM) theorem \cite{GutzwillerChaos}. If $0<\varepsilon \ll 1$,
the system is said to be \emph{quasi-integrable} and the KAM theorem
states that the tori are only slightly deformed if initially the system
is far from its resonances. A resonance is defined by $\sum _{n}{\ell _{n}}\omega _{n}=0$,
with ${\ell _{n}}$ integer. Note that the KAM-like structure of our
system depends only on the smallness of the parameter $\varepsilon $,
which can be modified by changing the properties of the potential.

\begin{figure}[htbp]
\includegraphics[  clip,
  angle=270,
  origin=lB]{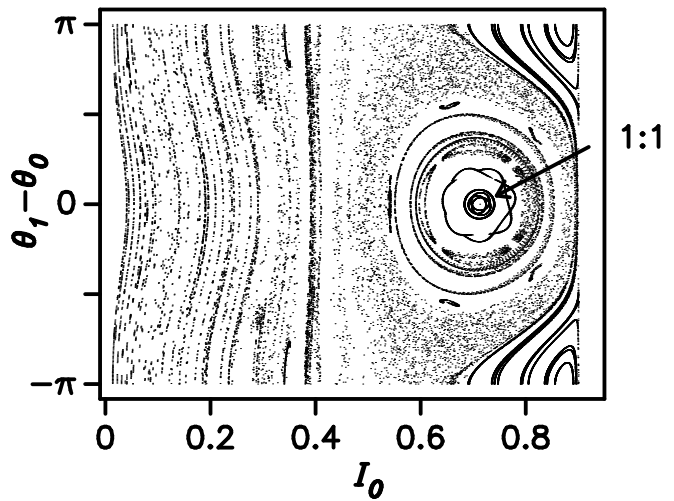}

\caption{\label{fig:Poincarre}Poincarré section of the generalized (quantum)
phase space, corresponding to the plane $(I_{0},\theta _{1}-\theta _{0})$.
The plane is defined by $I_{-1}=0.1$, $\theta _{-1}=\theta _{0}$,
all other angles and actions are zero, except $I_{1}=0.9-I_{0}$,
imposed by the normalization condition. One observes different types
of dynamics, ranging from regular to chaotic. The arrow indicates
the 1:1 resonance, and one can see around it various frequency-locking
signatures evidenced by the presence stability islands. Parameters
are the same as in Fig. \ref{fig:EvolutionGPE}.}
\end{figure}

In order to illustrate the above conclusions, consider the simple
case in which the BEC initially projects only on three adjacent WS
states, labeled -1,0,1. During the evolution, the spread of the wave
function to other WS states is very small %
\footnote{Population transfer between wells occurs if the nonlinear frequency
shift brings the energy levels of neighbor wells into degeneracy {[}cf.
Eq.(\ref{eq:ResCond}){]}. This condition cannot be satisfied for
wells with a low population.%
}. We performed a numerical integration of the equations of motion
and plotted a Poincarré section {[}of the generalized phase space
$(I,\theta )${]}, corresponding to the plane $(I_{0},\theta _{1}-\theta _{0})$,
displayed in Fig. \ref{fig:Poincarre}, which presents characteristic
features of a \emph{classically} chaotic system, despite the fact
the a BEC is a quantum object.

A principal resonance is observed when the ratio of two frequencies
is a rational number, e. g. $\omega _{n}/\omega _{m}=a/b$ ($a$ and
$b$ integers). The simplest case, $a=b=1$, called {}``1:1'' resonance,
corresponds to the resonance condition {[}cf. Eq.(\ref{eq:Freq}){]}\begin{equation}
\hbar \omega _{B}=U_{0}\chi _{000}\left(I_{0}-I_{1}\right).\label{eq:ResCond}\end{equation}
 for the two neighbor sites $n=0$, $m=1$. It has a simple interpretation:
the energy levels corresponding to the two wells become quasi-degenerate
due to the nonlinear shift, thus favoring population exchanges. Higher
order resonances affect smaller regions of the phase space. Thus,
in order to observe chaotic behavior, the initial wave function shall
not be too smoothly distributed on the lattice: if all $I_{n}$ are
almost identical, only higher order resonances will show up. The arrow
in Fig. \ref{fig:Poincarre} indicates the 1:1 resonance, which is
surrounded by other signatures of frequency-locking, characterized
by the presence of stability islands. The groups of three and five
stability islands correspond resp. to the resonances 3:1 and 5:1.
The quantum phase space has thus a KAM-like structure. Far from resonances
regular motion is displayed (left part of the Poincarré section).
In this case, the nonlinearity introduces population-dependent frequencies
due to the frequency pulling effect. The emergence of these frequencies
eventually blur Bloch oscillations (which would correspond to vertical
straight lines).

In conclusion, we have presented in this paper a quantum system that
displays a behavior very similar to \emph{classical} chaos, thanks
to the presence of a nonlinearity due to many-body interactions. As
we emphasized above, this is not \emph{quantum chao}s, at least in
the most widely accepted definition of the term. It is closer to classical
Hamiltonian chaos, except for the fact that it is observed in a generalized
phase space. The structure of the phase space can be interpreted in
the general frame of the KAM theorem. This leaves open many stimulating
questions: Are the conditions leading to Gross-Pitaevskii approach
essential for observing {}``classical'' chaos with a BEC, or the
same phenomenon can manifest itself in less restrictive conditions?
What are the necessary and sufficient conditions for observing {}``classical''
chaos with BECs? Can one find experimentally realizable situations
in which a quantum effect (e.g. quantum interference) coexists with
chaos, and how chaos will affect it? How {}``classical'' chaos with
BECs is affected by decoherence? These few questions suggest that
the present work opens an interesting way for investigations of the
quantum-classical limit in mesoscopic systems.

\begin{acknowledgments}
The authors are in debt to D. Delande for fruitful discussions. This
work is partially supported by a contract {}``ACI Photonique'' of
the Ministère de la Recherche. Laboratoire de Physique des Lasers,
Atomes et Mol\'{e}cules (PhLAM) is Unit\'{e} Mixte de Recherche
UMR 8523 du CNRS et de l'Universit\'{e} des Sciences et Technologies
de Lille. Centre d'Etudes et Recherches Lasers et Applications (CERLA)
is Fédération de Recherche FR 2416 du CNRS.
\end{acknowledgments}
\bibliographystyle{apsrev}
%\bibliography{/home/jcg/papers/ArtDataBase,/home/jcg/papers/Books}

\end{document}